\documentclass[acmtog]{acmart}

\usepackage{booktabs} 
\usepackage{enumitem}
\usepackage{multirow}
\usepackage{tabularx}
\usepackage{array}
\newcolumntype{Y}{>{\centering\arraybackslash}X}

\usepackage[table]{xcolor} 
\definecolor{tabfirst}{rgb}{1, 0.7, 0.7}  
\definecolor{tabsecond}{rgb}{1, 0.85, 0.7} 
\definecolor{tabthird}{rgb}{1, 1, 0.7}    

\setcopyright{cc}
\setcctype{by}
\acmJournal{TOG}
\acmYear{2026} \acmVolume{45} \acmNumber{6} \acmArticle{213}
\acmMonth{12} \acmDOI{10.1145/3842584}

\begin{document}
\title{TempFit: Panoptic Symbol Spotting via Template Fitting}

\author{Honghao Xu}
\orcid{0009-0005-6380-4078}
\affiliation{%
  \institution{Shenzhen University}
  \city{Shenzhen}
  \country{China}}
\email{littledaisy20001227@gmail.com}

\author{Hao Wang}
\orcid{0009-0001-0115-8358}
\affiliation{%
  \institution{SpeedBot Robotics Co., Ltd}
  \city{Changsha}
  \country{China}}
\email{21020100007@pop.zjgsu.edu.cn}

\author{Xiang Liu}
\orcid{0009-0002-8214-267X}
\affiliation{%
  \institution{Shenzhen University}
  \city{Shenzhen}
  \country{China}}
\email{lyr02302561@gmail.com}

\author{Juzhan Xu}
\orcid{0000-0002-5132-238X}
\affiliation{%
  \institution{Shenzhen University}
  \city{Shenzhen}
  \country{China}}
\email{juzhan.xu@gmail.com}

\author{Manfred Lau}
\orcid{0000-0001-9373-6620}
\affiliation{%
  \institution{Lakehead University}
  \city{Orillia}
  \country{Canada}}
\email{manfred.lau@gmail.com}

\author{Kai Xu}
\orcid{0000-0002-9054-0216}
\affiliation{%
  \institution{Institute of AI for Industries, Chinese Academy of Sciences and Jiangsu Key Laboratory of AI for Industries}
  \city{Nanjing}
  \country{China}}
\email{kevin.kai.xu@gmail.com}

\author{Ruizhen Hu}
\orcid{0000-0002-6798-0336}
\authornote{Corresponding author: Ruizhen Hu (ruizhen.hu@gmail.com)}
\affiliation{%
  \institution{Shenzhen University}
  \city{Shenzhen}
  \country{China}}
\email{ruizhen.hu@gmail.com}

\renewcommand{\shortauthors}{Honghao Xu, Hao Wang, Xiang Liu, Juzhan Xu, Manfred Lau, Kai Xu, and Ruizhen Hu}

%
%

%
%


\begin{teaserfigure}
    \includegraphics[width=\textwidth]{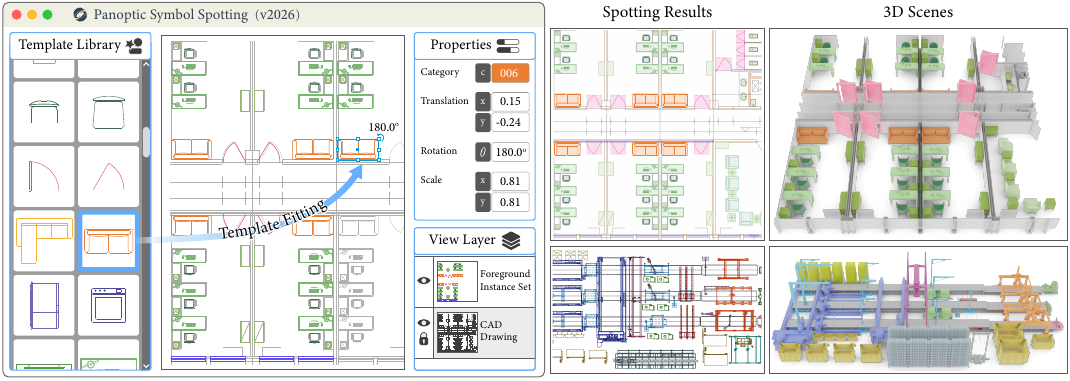}
    \Description{Motivated by the professional ``retrieve-and-transform'' CAD design workflow, we reformulate Panoptic Symbol Spotting as \emph{Template Fitting}. \textbf{Left:} given an input CAD drawing, our method retrieves the matching template from a predefined library and aligns it via pose prediction to produce the panoptic result. \textbf{Right:} since each template can be canonically associated with a 3D model, the detected symbols directly support downstream 3D applications such as digital twin reconstruction, illustrated here on floorplan (top) and industrial (bottom) scenes.}
    \caption{Motivated by the professional ``retrieve-and-transform'' CAD design workflow, we reformulate Panoptic Symbol Spotting as \emph{Template Fitting}. \textbf{Left:} given an input CAD drawing, our method retrieves the matching template from a predefined library and aligns it via pose prediction to produce the panoptic result. \textbf{Right:} since each template can be canonically associated with a 3D model, the detected symbols directly support downstream 3D applications such as digital twin reconstruction, illustrated here on floorplan (top) and industrial (bottom) scenes.}    \label{fig:teaser}
\end{teaserfigure}

\begin{abstract}
In this paper, we introduce \emph{TempFit}, a novel approach for the problem of Panoptic Symbol Spotting on CAD drawings. Existing methods treat symbol spotting as a per-primitive classification or mask-prediction task, with no explicit constraint enforcing that the primitives assigned to one instance form a spatially connected group. As a result, predictions may erroneously merge primitives from distant symbols into a single instance or fragment a coherent symbol across multiple predictions. Motivated by the professional ``retrieve-and-transform'' CAD design workflow, we reformulate this task as a \emph{Template Fitting} problem, where each foreground instance is detected by retrieving and aligning a canonical template from a library, enforcing structural completeness by construction. Building on this formulation, we present an autoregressive transformer that unifies template retrieval and pose prediction within a single end-to-end pipeline. The framework treats the template library as an explicit geometric memory, parsing background regions in parallel via learnable background tokens and synthesizing foreground instances autoregressively while attending to template-specific geometry at each step. Experiments on FloorplanCAD and our newly collected IndustryCAD dataset, a more challenging benchmark with complex industrial symbols, heavy occlusion, and wide scale variation, demonstrate state-of-the-art performance with particularly large gains on IndustryCAD. Project page: \url{https://tempfit-project.github.io/}.
\end{abstract}

\maketitle

\section{Introduction}

Panoptic symbol spotting \cite{fan2021floorplancad} aims to simultaneously detect and segment countable foreground object instances (e.g., furniture in a floorplan) while parsing uncountable background 
regions (e.g., walls) from 2D CAD drawings.
In architecture, engineering and construction (AEC) industries, this task underpins critical downstream applications such as building information modeling (BIM) and 3D digital twin reconstruction, where accurate instance-level parsing is essential for faithfully recovering the structural semantics from the given CAD drawings.
However, this problem is highly non-trivial, as real-world drawings often suffer from severe clutter, symbol occlusions, and significant variations in appearance and structure within the same category\cite{fan2022cadtransformer}.

To tackle this challenging task, existing methods can be broadly categorized into two paradigms: primitive-wise and instance-wise methods. Primitive-wise methods \cite{fan2021floorplancad, fan2022cadtransformer, zheng2022gat} treat each graphical primitive as an independent unit, employing GNNs or Transformers to predict a semantic label for each primitive, followed by heuristic grouping techniques to assemble them into object instances. 
Instance-wise methods \cite{liu2024symbol, liu2024sympointv2, yang2024cadspotting, wei2025point, luo2025archcad} adopt a query-based segmentation prediction paradigm, where each query outputs a class label and a binary mask indicating which primitives belong to that instance. 
Despite their differences, both paradigms share a fundamental limitation: they treat symbol spotting as a per-primitive classification or mask-prediction problem, without explicitly modelling the local spatial topology within each symbol---such as the connectivity of lines and arcs that compose a chair or a door. Consequently, no constraint enforces that the primitives assigned to one instance form a spatially connected group, and predictions may erroneously merge primitives from distant symbols into a single instance or fragment a coherent symbol across multiple predictions.

To overcome this spatial incoherence, we reformulate the task as a \emph{Template Fitting} problem, motivated by the ``Retrieve-and-Transform'' workflow of professional CAD design: designers maintain a library of canonical templates and compose drawings by retrieving templates and adjusting their poses, rather than drawing stroke-by-stroke. Following this paradigm, each foreground symbol is modeled as the instantiation of a template under a 2D transformation---the model retrieves the best-matching template and predicts the transformation parameters that align it with the target region. Since every prediction is, by construction, a valid template, its primitives automatically form a structurally complete and topologically coherent group, directly eliminating the merge and fragmentation errors common in prior paradigms. Each template can also be canonically associated with a 3D model, enabling detected symbols to directly serve downstream BIM and digital twin reconstruction, as shown in Figure \ref{fig:teaser}. We argue that this Template Fitting formulation represents a more principled approach to Panoptic Symbol Spotting, as it is fundamentally aligned with how CAD drawings are originally authored.

While the Template Fitting formulation is conceptually appealing, executing it effectively for 2D CAD symbol spotting is non-trivial. This formulation has been explored in two adjacent settings, but neither solution transfers directly to our problem. 
The first is geometric shape abstraction \cite{ye2025primitiveanything,huang2023learning, xu2023shapegeneration}, which approximates a 3D shape with a set of parameterized templates such as cuboids or cylinders. In this setting the template library is small and fixed, so existing methods typically memorize template geometry implicitly within the network weights, rather than treating it as an explicit input. This design prevents generalization to unseen templates and is unsuitable for our task, where the template library is large and the shapes are too complex to be reliably memorized by a network.
The second is scene CAD recomposition \cite{avetisyan2019scan2cad,kuo2020mask2cad, li2023autoscan2cad, leng2024shape}, which aligns observed objects in a 3D scan or RGB image with canonical CAD meshes from an external library. This setting is closer to ours: the template library is large and complex, and its geometry is provided as explicit input. However, these methods adopt a staged detect-then-fit pipeline: an object detector first localizes each instance based on category-level appearance alone, and the template retrieval and pose-fitting steps then operate on each detected region independently. This decoupling causes 
per-instance error propagation ---a missed or misaligned detection constrains the downstream fitting result for the same instance.


To address these challenges, we introduce \emph{TempFit}, an autoregressive transformer framework that unifies template retrieval and pose prediction into a single, end-to-end process. The core idea is to treat the template library as an explicit geometric memory: each template is encoded into a sequence of tokens and prepended to the input, enabling the model to attend to detailed shape geometry throughout the fitting process. The scene is processed through two components. Background regions are parsed in parallel via learnable background tokens, each predicting a category mask for a specific region type. Foreground instances are generated autoregressively: at each step, the model attends over the encoded template memory to jointly select and align a template with its target region in the drawing. This design directly addresses the limitations of prior Template Fitting approaches. 
This design avoids memorizing geometry in network weights (unlike shape abstraction) and eliminates the error accumulation of staged pipelines (unlike scene CAD recomposition).



To validate the effectiveness of our proposed method, we conducted experiments on the FloorPlanCAD \cite{fan2021floorplancad} dataset, where our method achieves significant improvements over state-of-the-art approaches. Furthermore, we collected IndustryCAD, a new industrial CAD dataset that is more challenging than FloorPlanCAD in three aspects: greater symbol complexity, more severe occlusion, and wider scale variation. On this dataset, our method yields even larger gains over prior methods. 



Our main contributions can be summarized as follows:
\begin{itemize}[leftmargin=*]
\item We propose a novel perspective that reformulates Panoptic Symbol Spotting as a \emph{Template Fitting} task, inherently enforcing structural completeness and topological consistency for each symbol, aligned with how CAD drawings are authored.
\item We introduce \emph{TempFit}, an autoregressive transformer that treats the template library as an explicit geometric memory and unifies template retrieval with pose prediction into a single, end-to-end process, eliminating the error accumulation of staged pipelines while generalizing to large and extensible template libraries.
\item We conduct extensive experiments on FloorPlanCAD and our newly collected IndustryCAD dataset, achieving state-of-the-art results on both benchmarks with particularly large gains on the more challenging IndustryCAD.
\end{itemize}

\section{Related Work}

Our work intersects three lines of research: Panoptic Symbol Spotting, the target task we aim to improve; 3D Shape Abstraction, which fits parametric templates to approximate geometry; and Scene CAD Recomposition, which aligns canonical CAD meshes with visual observations. We review each and position TempFit accordingly.

\subsection{Panoptic Symbol Spotting}
The Panoptic Symbol Spotting task was introduced by \cite{fan2021floorplancad} along with the FloorPlanCAD benchmark, requiring the joint parsing of countable instances (e.g., furniture) and uncountable semantic regions (e.g., walls) from CAD drawings. Existing approaches fall into two paradigms: Primitive-wise and Instance-wise methods.

\paragraph{Primitive-wise Methods}
Early approaches typically treat each graphical primitive as an independent processing unit. The core paradigm involves using Graph Neural Networks (GNNs) or Transformers to predict a semantic label for each primitive individually, followed by heuristic grouping techniques (e.g., bounding box detection or clustering) to assemble them into instances. For example, PanCADNet \cite{fan2021floorplancad} combines graph convolutional networks \cite{jiang2019semi} for semantic segmentation with Faster R-CNN \cite{ren2015faster} for instance grouping. GAT-CADNet \cite{zheng2022gat} employs Graph Attention Networks to capture topological relations between primitives, while CADTransformer \cite{fan2022cadtransformer} utilizes Vision Transformers \cite{dosovitskiy2020image} on rasterized inputs to aggregate primitive features.

\paragraph{Instance-wise Methods}
Following the success of DETR \cite{carion2020end}, the field has shifted towards query-based set prediction. These methods formulate the task as predicting a set of instance masks, where each query outputs a semantic class and a binary mask $\in \{0,1\}^N$ indicating which primitives belong to that instance. Based on their input representation, these methods can be categorized into three groups: (1) Point-based: SymPoint \cite{liu2024symbol} and SymPoint-V2 \cite{liu2024sympointv2} treat each primitive as a single point;
(2) Dense-point-based: CADSpotting \cite{yang2024cadspotting} densely samples points along primitives to learn richer features; (3) Line-based: VecFormer \cite{wei2025point} directly processes vector segments to better preserve geometric fidelity.

Despite their differences, both primitive-wise and instance-wise paradigms share a fundamental limitation: they treat symbol spotting as a per-primitive classification or mask-prediction task, with no explicit constraint enforcing that the primitives assigned to one instance form a spatially connected group. Consequently, predictions may erroneously merge primitives from distant symbols into a single instance, or fragment a coherent symbol across multiple predictions. TempFit reformulates the task as template fitting, where each predicted instance is by construction an instantiation of a valid template whose primitives form a topologically coherent group.

\subsection{3D Shape Abstraction}
Shape Abstraction aims to approximate complex geometry using a set of parameterized primitives. Existing work spans a spectrum of geometric complexity. Simple forms include cuboids~\cite{gupta2010blocks,li2017grass,mo2019structurenet,tulsiani2017learning} and cylinders~\cite{binford1975visual}. More expressive representations include superquadrics~\cite{paschalidou2019superquadrics,pentland1986parts}, convex polytopes~\cite{chen2020bsp,deng2020cvxnet}, learned implicit fields~\cite{chen2019bae,gadelha2020learning,genova2019learning}, and B-spline patches~\cite{sharma2020parsenet}.

These methods broadly fall into two paradigms: (1) Optimization-based methods directly optimize parameters to minimize reconstruction error \cite{liu2023marching, pitas2002morphological, monnier2023differentiable}. While mathematically precise, they often fragment semantic objects into incoherent pieces as they lack semantic awareness. (2) Learning-based methods employ neural networks to predict primitive parameters directly, trained via supervised \cite{li2017grass, jones2020shapeassembly, mo2019structurenet, ye2025primitiveanything, li2019spfn} or unsupervised \cite{chen2020bsp, deng2020cvxnet, genova2020local, sun2019learning, chen2024daenet} reconstruction objectives.

Applying these methods to CAD symbol spotting is fundamentally limited by their reliance on a small, fixed primitive library. Because the library is small, existing methods memorize primitive geometry implicitly within the network weights rather than treating it as an explicit input. This implicit encoding fails to generalize to our setting, where the template library is much larger and the symbols are too topologically complex to be reliably memorized by a network. A recent work \cite{xu2023partassembly} takes an explicit part library as input but relies on per-shape iterative optimization that does not scale to scene-level decoding. In contrast, TempFit explicitly encodes template geometry as input, supports an extensible vocabulary via dynamic matching, and produces all instances in a single autoregressive pass without test-time optimization.

\begin{figure*}[!t]
    \centering
    \includegraphics[width=\textwidth]{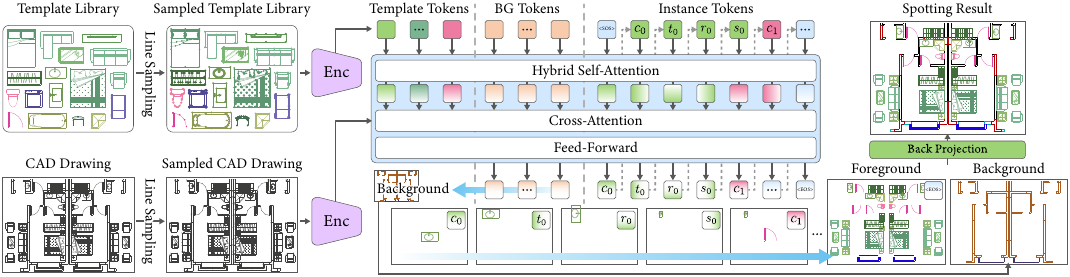}
    \Description{Overview of TempFit. \textbf{Left}: The input drawing and canonical templates are discretized into line segments and encoded by a shared backbone, yielding drawing tokens and template tokens. \textbf{Middle}: We prepend the template tokens as an explicit geometric memory and process the scene via Background Segmentation and Foreground Template Fitting: the former predicts per-category masks in parallel via learnable background tokens, while the latter generates instances autoregressively by retrieving a template and predicting its transformation token-by-token. \textbf{Right}: The predicted background masks and foreground templates are projected back onto the input primitives to produce the final panoptic result.}
    \caption{Overview of TempFit. \textbf{Left}: The input drawing and canonical templates are discretized into line segments and encoded by a shared backbone, yielding drawing tokens and template tokens. \textbf{Middle}: We prepend the template tokens as an explicit geometric memory and process the scene via Background Segmentation and Foreground Template Fitting: the former predicts per-category masks in parallel via learnable background tokens, while the latter generates instances autoregressively by retrieving a template and predicting its transformation token-by-token. \textbf{Right}: The predicted background masks and foreground templates are projected back onto the input primitives to produce the final panoptic result.}
    \label{fig:network}
\end{figure*}

\subsection{Scene CAD Recomposition}
Scene CAD recomposition aligns observed objects in 3D scans or RGB images with canonical CAD meshes from external libraries such as ShapeNet~\cite{chang2015shapenet}. This setting is closest to ours, since the template library is large and provided as explicit input. However, existing methods adopt a \emph{detect-then-fit} pipeline.

Representative scan-based methods include Scan2CAD~\cite{avetisyan2019scan2cad}, which aligns CAD via heatmap correspondences between scan keypoints and CAD voxels; SceneCAD~\cite{avetisyan2020scenecad}, which models object--object and object--layout relations with a GNN; and AutoScan2CAD~\cite{li2023autoscan2cad}, which combines relation-guided global optimization with active scanning. Image-based methods include Mask2CAD~\cite{kuo2020mask2cad}, which learns an image--CAD embedding for retrieval after detection; Patch2CAD~\cite{kuo2021patch2cad}, which uses patch-wise embeddings for part-level matching when exact CAD models are unavailable; ROCA~\cite{gumeli2022roca}, which predicts depth and normalized coordinates for differentiable pose solving; and DiffCAD~\cite{gao2024diffcad}, which learns a weakly-supervised diffusion cascade over scale, pose, and shape.

The critical weakness of this decoupled design is error accumulation: detection failures---missed objects, misaligned boxes---propagate irreversibly to the fitting stage with no recovery path. The downstream retrieval and pose estimation are upper-bounded by the quality of the initial proposals, and information that would aid detection, such as template-specific geometry, is unavailable to the detector. In contrast, TempFit unifies template retrieval and transformation prediction within a single end-to-end transformer, allowing template geometry to directly inform instance detection and eliminating the intermediate detection bottleneck entirely.

\section{Method}

\label{sec:problem_formulation}
Given a CAD drawing represented as an unordered set of graphical primitives $\mathcal{E} = \{e_k\}_{k=1}^K$,  the goal of panoptic symbol spotting is to learn a mapping $\mathcal{F}: e_k \mapsto (l_k, z_k)$. Here, $l_k \in \mathcal{L}$ denotes a semantic category, and $z_k \in \mathbb{N}$ is an instance identifier. The category set $\mathcal{L}$ consists of background classes $\mathcal{L}_{\text{bg}}$ and foreground classes $\mathcal{L}_{\text{fg}}$, which we handle through distinct mechanisms within a unified pipeline, as illustrated in Figure~\ref{fig:network}.

Unlike prior approaches that adopt a unified segmentation-based paradigm for both foreground and background objects, we handle them separately. For background regions (e.g., walls, floors), we follow the standard segmentation formulation: the model outputs a per-category binary mask over primitives, as in prior work. For foreground instances, we instead adopt a Template Fitting paradigm: each instance is detected by retrieving a canonical template from a predefined library $\mathcal{T}$ and predicting its 2D pose. Section~\ref{sec:TempFit} formalizes the approach and Section~\ref{sec:network} details the network architecture.

Finally, we unify both outputs by projecting them back onto the primitive space: background masks directly assign semantic labels to their covered primitives, while each fitted foreground template is instantiated on the canvas and matched to nearby primitives via geometric alignment. This recovers the full panoptic mapping $\mathcal{F}: e_k \mapsto (l_k, z_k)$ for every primitive $e_k$. The back projection procedure is detailed in Section~\ref{sec:back_projection}.

\subsection{Template Fitting}
\label{sec:TempFit}

\paragraph{Problem Formulation}
We now formalize the template fitting approach described above. The target panoptic output decomposes into two sets $\mathcal{O} = \mathcal{O}_{\text{bg}} \cup \mathcal{O}_{\text{fg}}$. For background, each category is represented by a binary mask over primitives:
$\mathcal{O}_{\text{bg}} = \{ (c_j, \mathbf{m}_j) \mid c_j \in \mathcal{L}_{\text{bg}} \}$, where $\mathbf{m}_j \in \{0, 1\}^K$ indicates which primitives belong to category $c_j$.
For foreground, each instance is parameterized as a retrieved template under a 2D transformation:
\begin{equation}
\label{eq:ofg}
\mathcal{O}_{\text{fg}} = \left\{ (c_i, \mathbf{t}_i, \theta_i, \mathbf{s}_i) \mid i = 1, \dots, N, \ c_i \in \mathcal{T} \right\},
\end{equation}
where $c_i$ indexes a template prototype in $\mathcal{T}$, and $(\mathbf{t}_i,\theta_i,\mathbf{s}_i)$ denotes its 2D translation, rotation, and anisotropic scale.
We formulate panoptic parsing as a joint conditional generation process by factorizing:
\begin{equation}
\label{eq:formulation}
P(\mathcal{O} \mid \mathcal{E}, \mathcal{T}) = P(\mathcal{O}_{\text{bg}} \mid \mathcal{E}, \mathcal{T}) \cdot \prod_{i=1}^{N} P(o_i \mid \mathcal{E}, \mathcal{T}, o_{<i}),
\end{equation}
where the first term predicts background regions and the second term generates foreground instances.

The template library $\mathcal{T}$ can be sourced flexibly depending on the application. In industrial settings, domain experts often maintain 3D asset libraries of the equipment present at a site; templates are then obtained by projecting these canonical models to 2D (as in IndustryCAD, cf.\ Section~\ref{sec:industrycad}). When no such library is available, it can be automatically constructed by clustering training-set instances (as we do for FloorPlanCAD, cf.\ Section~\ref{sec:datasets}).

\paragraph{Framework Design}
Inspired by shape abstraction methods that decompose a target into a set of parameterized templates~\cite{ye2025primitiveanything}, we adopt an autoregressive architecture for the foreground term: instances are generated sequentially, each conditioned on previously generated ones. This mirrors the retrieve-then-transform workflow of CAD designers, where each symbol is individually selected and placed.

A key limitation of prior shape abstraction methods is that they memorize template geometry implicitly within network weights, preventing generalization to large or extensible template libraries. To address this, we encode the template library $\mathcal{T}$ as explicit input, allowing the model to access template-specific geometry through attention rather than abstract memorization. This design draws from prefix-tuning~\cite{li2021prefix} in language modeling, with encoded templates replacing learnable virtual prompts.

For background regions, each category requires only a single segmentation mask and no sequential dependency across categories. Autoregressive generation is therefore unnecessary. We instead predict all background masks in parallel \emph{before} the foreground autoregressive generation begins, so the resulting global semantic layout serves as contextual cues for subsequent instance prediction. This informs the model, for instance, where walls and open spaces lie before it places doors and furniture, helping resolve ambiguities in cluttered regions.

These three design choices---autoregressive instance generation, explicit geometric memory, and background-first prediction---are unified through a prefix attention mechanism, detailed in Section~\ref{sec:network}.

\subsection{Network Architecture}
\label{sec:network}

\paragraph{CAD and Template Encoding}
We adopt the line-based representation from VecFormer~\cite{wei2025point} and encode both the input drawing and the template library through a shared pipeline. All graphical primitives are discretized into line segments via uniform sampling, each assigned a geometric feature vector (length, direction, midpoint, and primitive centroid). These line segments, 
together with their geometric features, 
pass through a shared PointTransformerV3~\cite{wu2024point} backbone. For the drawing, line tokens are aggregated into primitive-level tokens $\mathbf{T}_{\text{draw}}\in \mathbb{R}^{K \times d}$ via summed max- and average-pooling. For each template, all line tokens are mean-pooled into a single token, forming the explicit geometric memory $\mathbf{T}_{\mathrm{tpl}} \in \mathbb{R}^{|\mathcal{T}| \times d}$.

\paragraph{Background Segmentation}
To estimate the first term in Equation~\ref{eq:formulation}, we formulate background segmentation as a parallel detection task. $\mathbf{T}_{\text{bg}}\in \mathbb{R}^{|\mathcal{L}_{\text{bg}}| \times d}$ is a set of learnable embeddings, 
each fixed to a specific background category, 
randomly initialized and optimized end-to-end. 
Unlike category-agnostic DETR queries dynamically assigned to targets via bipartite matching, these tokens preserve their category-specific semantics across drawings.
During the forward pass, these tokens interact with the drawing context $\mathbf{T}_{\text{draw}}$ via cross-attention. Let $\mathbf{h}_{\text{bg}}^j \in \mathbb{R}^d$ denote the output token for the $j$-th background category. Two parallel heads decode it: an existence classifier and a mask prediction head:
\begin{equation}
p_j = \sigma\big(\text{MLP}_{\text{cls}}(\mathbf{h}_{\text{bg}}^j)\big),
\quad
\mathbf{m}_j = \sigma\big(\text{MLP}_{\text{draw}}(\mathbf{T}_{\text{draw}}) \cdot \text{MLP}_{\text{mask}}(\mathbf{h}_{\text{bg}}^j)^\top\big),
\end{equation}
where $\mathbf{m}_j \in [0,1]^K$ is the per-primitive membership score for category $c_j$. Together, $\{p_j\}$ and $\{\mathbf{m}_j\}$ capture the scene layout via per-category existence scores and segmentation masks.

\begin{figure}[!t]
    \centering
    \includegraphics[width=\columnwidth]{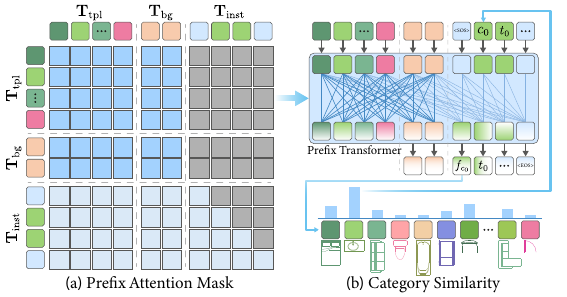}
    \Description{\textbf{(a) Prefix attention mask:} bidirectional attention within $[\mathbf{T}_\text{tpl}, \mathbf{T}_\text{bg}]$ and causal masking on $\mathbf{T}_\text{inst}$ (gray cells block attention). \textbf{(b) Class prediction and feature injection:} cosine similarity matching against normalized template tokens, followed by injecting the top-scoring template token into the sequence to condition pose estimation.
    }
    \caption{\textbf{(a) Prefix attention mask:} bidirectional attention within $[\mathbf{T}_\text{tpl}, \mathbf{T}_\text{bg}]$ and causal masking on $\mathbf{T}_\text{inst}$ (gray cells block attention). \textbf{(b) Class prediction and feature injection:} cosine similarity matching against normalized template tokens, followed by injecting the top-scoring template token into the sequence to condition pose estimation.
    }
    \label{fig:attention_mask}
\end{figure}

\begin{figure*}[!t]
    \centering
    \includegraphics[width=\linewidth]{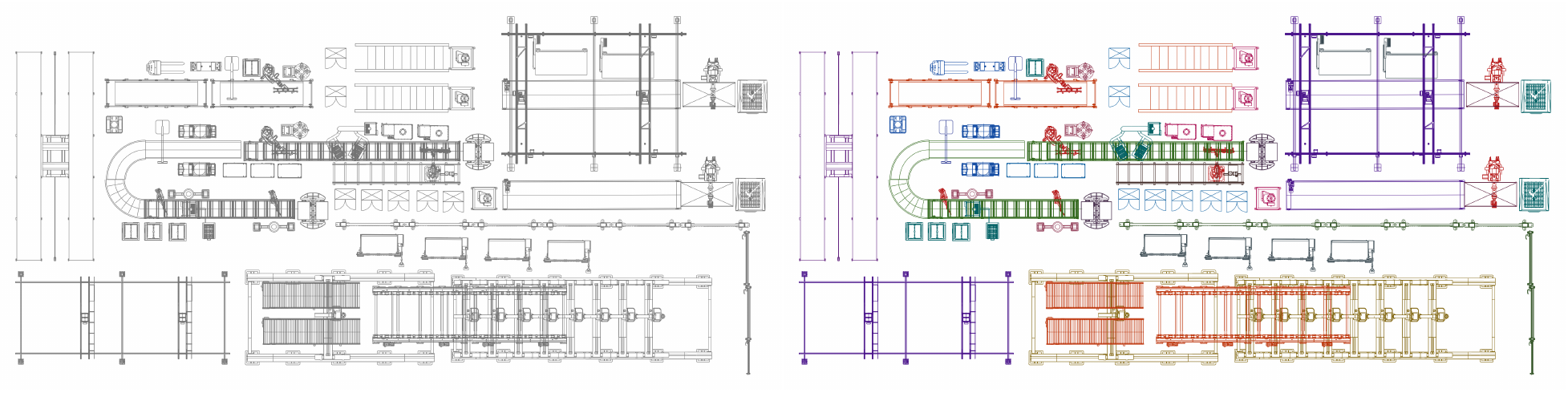}
    \Description{An annotated example from IndustryCAD. \textbf{Left}: The raw factory drawing with text removed. \textbf{Right}: Per-primitive ground-truth annotations across 23 equipment categories, each shown in a distinct color.}
    \caption{An annotated example from IndustryCAD. \textbf{Left}: The raw factory drawing with text removed. \textbf{Right}: Per-primitive ground-truth annotations across 23 equipment categories, each shown in a distinct color.}
    \label{fig:industry_snapshot}
\end{figure*}

\paragraph{Foreground Template Fitting}
We realize the second term in Equation~\ref{eq:formulation} as an attribute-wise autoregressive process. As illustrated in Figure~\ref{fig:attention_mask}(a), the input sequence is $[\mathbf{T}_{\mathrm{tpl}}, \mathbf{T}_{\text{bg}}, \mathbf{T}_{\mathrm{inst}}]$, where $\mathbf{T}_{\mathrm{inst}}$ denotes the autoregressively constructed sequence of instance tokens. We enable bidirectional attention within the prefix $[\mathbf{T}_{\mathrm{tpl}}, \mathbf{T}_{\text{bg}}]$ and causal masking on $\mathbf{T}_{\mathrm{inst}}$, while instance tokens attend to the full prefix to condition each prediction on both the geometric library and the predicted background layout. Let $\mathcal{H}_i = \{ \mathbf{T}_{\text{draw}}, \mathbf{T}_{\mathrm{tpl}}, \mathbf{T}_{\text{bg}}, o_{<i} \}$ denote the context history. We decompose:
\begin{equation}
P(o_i \mid \mathcal{H}_i) = \underbrace{P(c_i \mid \mathcal{H}_i)}_{\text{Retrieve}} \cdot \underbrace{P(\mathbf{t}_i, \theta_i, \mathbf{s}_i \mid c_i, \mathcal{H}_i)}_{\text{Transform}}.
\end{equation}

\emph{Retrieve.} Instead of a fixed classifier, we perform dynamic metric matching against the template prefix. The normalized output embedding $\mathbf{f}_{\text{cls}}$ is matched via cosine similarity against all normalized template tokens:
\begin{equation}
P(c_i \mid \mathbf{f}_{\text{cls}}, \mathbf{T}_\text{tpl}) = \text{Softmax}\left( \mathbf{f}_{\text{cls}} \cdot \mathbf{T}_\text{tpl}^\top \right).
\end{equation}
The category with maximal probability is selected. Crucially, as shown in Figure~\ref{fig:attention_mask}(b), we reuse the selected template token $\mathbf{T}_{\mathrm{tpl}}[c_i]$ to initialize the next attribute input, explicitly conditioning pose estimation on the geometry of the retrieved prototype.

\emph{Transform.}
Conditioned on the injected geometric prior, the model sequentially predicts the transformation attributes $(\mathbf{t}_i,\theta_i,\mathbf{s}_i)$. Each attribute is discretized into $K$ uniform bins over its valid range and predicted as a categorical distribution. 
Each attribute generation attends to the drawing context, the template prefix, and previously predicted attributes, ensuring geometrically consistent placement.

\paragraph{Loss Function}
We train the Transformer by minimizing:
\begin{equation}
\label{eq:loss}
\mathcal{L} = \underbrace{\mathcal{L}_{\mathrm{cls}} + \mathcal{L}_{\mathrm{geo}} +  \mathcal{L}_{\mathrm{huber}}}_{\text{Foreground Instance Fitting}} + \underbrace{ \mathcal{L}_{\mathrm{valid}} + \mathcal{L}_{\mathrm{mask}}}_{\text{Background Segmentation}} + \underbrace{ \mathcal{L}_{\mathrm{recon}} + \mathcal{L}_{\mathrm{eos}}}_{\text{Auxiliary}},
\end{equation}
where $\mathcal{L}_{\mathrm{cls}}$ is a hierarchy-aware cross-entropy loss for template classification following~\cite{bertinetto2020making}, which penalizes intra-category confusions less than inter-category errors;
$\mathcal{L}_{\mathrm{geo}}$ supervises discrete transformation tokens $(t_i,\theta_i,s_i)$ via cross-entropy;
$\mathcal{L}_{\mathrm{huber}}$ applies Huber regression to continuous attributes obtained as expectations over discrete predictions;
$\mathcal{L}_{\mathrm{valid}}$ is a BCE loss for background category existence;
$\mathcal{L}_{\mathrm{mask}}$ combines BCE and Dice losses for background masks;
$\mathcal{L}_{\mathrm{recon}}$ is a Chamfer loss between sampled points of transformed templates and ground-truth instances;
and $\mathcal{L}_{\mathrm{eos}}$ supervises the stop-token probability via BCE.
All loss terms are summed with equal weights. Detailed formulations are provided in the supplementary material.

\subsection{Back Projection}
\label{sec:back_projection}
We synthesize the final spotting results by projecting both background and foreground predictions onto the primitive set: we first assign high-confidence primitives in the predicted background masks to their respective background categories, and then solve a robust primitive-to-instance assignment for the remaining primitives.

For each predicted instance $o_i$, we restrict candidates to a local set $\mathcal{E}_{\mathrm{loc}}$ consisting of unassigned primitives inside its inflated oriented bounding box.
For $e_k \in \mathcal{E}_{\mathrm{loc}}$, we define an alignment cost:
\begin{equation}
E(e_k,o_i) = \lambda_{\mathrm{ang}} E_{\mathrm{ang}} + \lambda_{\mathrm{dist}} E_{\mathrm{dist}} + \lambda_{\mathrm{sub}} E_{\mathrm{sub}},
\end{equation}
where $E_{\mathrm{ang}}$ penalizes orientation mismatch to the nearest instance primitive, $E_{\mathrm{dist}}$ measures the perpendicular distance, and $E_{\mathrm{sub}}$ penalizes projection overshoot beyond the bounds of $o_i$ to avoid semantic leakage (e.g., assigning a long wall segment to a short window). In overlapping regions, each primitive is assigned to the instance with the minimum cost.
Detailed formulations of the three cost terms and their parameter values are provided in the supplementary material.


\section{Experiments}

We evaluate on FloorPlanCAD and the newly introduced IndustryCAD. We first introduce the datasets~(Section~\ref{sec:datasets}) and describe the experimental setup~(Section~\ref{sec:experiment_settings}), then report results on FloorPlanCAD~(Section~\ref{sec:floorplan_results}) and IndustryCAD~(Section~\ref{sec:industrycad_results}), followed by ablation studies~(Section~\ref{sec:ablation}) verifying our core design choices and the effect of instance ordering, and a robustness analysis~(Section~\ref{sec:robustness}) examining autoregressive error accumulation, template library size, and cross-dataset transfer.

\begin{table*}[t]
\centering
\caption{Quantitative comparison on FloorplanCAD test set. The best and second-best results are highlighted by color. ''-'' denotes entries that cannot be reproduced under our protocol due to missing public implementations. Speed is single-sample inference latency on a single RTX 4090. Under the Hungarian metric, our method achieves the highest PQ on both foreground and background categories.
}
\label{tab:floorplancad}

\begin{tabularx}{\textwidth}{l c c *{6}{Y}}
\toprule
\multirow{2}{*}{Methods}
& \multirow{2}{*}{\#Params}
& \multirow{2}{*}{Speed}
& \multicolumn{3}{c}{Greedy Metric}
& \multicolumn{3}{c}{Hungarian Metric} \\
\cmidrule(lr){4-6} \cmidrule(lr){7-9}
&
&
& PQ
& $\mathrm{PQ}_{\mathrm{fg}}$
& $\mathrm{PQ}_{\mathrm{bg}}$
& PQ
& $\mathrm{PQ}_{\mathrm{fg}}$
& $\mathrm{PQ}_{\mathrm{bg}}$ \\
\midrule

PanCADNet \cite{fan2021floorplancad}
& -- & --
& 59.5 & 65.6 & 58.7
& --   & --   & -- \\

CADTransformer \cite{fan2022cadtransformer}
& -- & --
& 68.9 & 78.5 & 58.6
& --   & --   & -- \\

GAT-CADNet \cite{zheng2022gat}
& -- & --
& 73.7 & --   & --
& --   & --   & -- \\
\midrule

SymPoint \cite{liu2024symbol}
& \hspace{1pt} 35.1 M & \hspace{1pt} 84.7 ms
& 83.3 & 84.1 & 48.2
& 61.7 & 64.8 & 29.0 \\

SymPointV2 \cite{liu2024sympointv2}
& \hspace{1pt} 35.1 M & 113.3 ms
& \cellcolor{tabsecond}90.1
& \cellcolor{tabsecond}90.8
& \cellcolor{tabsecond}80.8
& \cellcolor{tabsecond}77.0
& \cellcolor{tabsecond}78.9
& 58.0 \\

CADSpotting \cite{yang2024cadspotting}
& \hspace{1pt}  49.5 M & \hspace{1pt} 42.5 ms
& 88.9 & 89.7 & 80.6
& --   & --   & -- \\

VecFormer \cite{wei2025point}
& \hspace{1pt} 54.2 M & \hspace{1pt}  32.9 ms
& \cellcolor{tabfirst}91.1
& \cellcolor{tabfirst}91.8
& \cellcolor{tabfirst}90.4
& 48.9 & 59.4 & 18.7 \\
\midrule

PrimitiveAnything \cite{ye2025primitiveanything}
& 104.6 M & 201.3 ms
& 78.3 & 81.3 & 75.2
& 71.4 & 73.9
& \cellcolor{tabsecond}61.7 \\

DEIM+ICP \cite{huang2025deim}
& \hspace{1pt} 62.0 M & \hspace{1pt} 65.9 ms
& 82.0 & 82.0 & --
& 74.4 & 74.4 & -- \\
\midrule

Ours
& 108.3 M & 279.0 ms
& 86.2 & 86.8 & 79.6
& \cellcolor{tabfirst}80.1
& \cellcolor{tabfirst}81.7
& \cellcolor{tabfirst}65.4 \\
\bottomrule
\end{tabularx}
\end{table*}

\subsection{Datasets}
\label{sec:datasets}
We evaluate on two complementary benchmarks: FloorPlanCAD~\cite{fan2021floorplancad} and our newly introduced IndustryCAD, a real-world industrial dataset characterized by high primitive complexity, severe inter-instance occlusion, and extreme scale variation.

\paragraph{FloorPlanCAD} We follow the standard split (6{,}965 / 810 / 3{,}827 train/val/test). Since the dataset does not ship with an explicit template library, we construct one from the training set.

We first compute a rotation- and scale-invariant shape-context descriptor~\cite{nguyen2008symbol} for each of the 103{,}959 training instances and evaluate pairwise similarity within each category. We set a similarity threshold at the top 1\% of the pairwise similarity distribution, so that only the most similar 1\% of instance pairs are treated as geometrically equivalent. For each instance, we count how many others exceed this threshold, yielding a repetition count that measures how frequently its geometry recurs in the training set. We then construct the template library by greedy coverage: starting with the highest-count instance, we add it to the library and simultaneously remove it and all instances whose similarity to it exceeds the threshold. We repeat this on the remaining set until at least 95\% of all instances have been removed. This procedure yields 462 canonical prototypes, which form the template library used throughout training and inference.

\paragraph{IndustryCAD}\label{sec:industrycad}
Unlike existing CAD benchmarks that focus on residential and commercial buildings, IndustryCAD is collected from real factory scenes across 71 distinct industrial sites. We design 23 equipment categories covering common industrial machinery and adopt the per-primitive annotation protocol of FloorPlanCAD; the dataset contains 34{,}091 annotated instances in total. One example is shown in Figure~\ref{fig:industry_snapshot}. Each industrial drawing covers a large physical area (mean 3{,}779.9~m$^2$), so we partition every drawing into fixed 30~m~$\times$~30~m blocks, yielding 2{,}745 blocks. We randomly split the 71 scenes into 63 for training and 8 for testing (2{,}213 / 532 blocks and 28{,}153 / 5{,}938 instances). The template library consists of 76 prototypes projected from 3D asset libraries at the data source.

IndustryCAD presents three key challenges that distinguish it from existing benchmarks. (a)~\textit{Complexity:} industrial equipment is composed of far more geometric primitives than furniture symbols, 2{,}233 on average per instance, versus 20.3 in FloorPlanCAD, substantially increasing the difficulty of instance grouping and semantic recognition. (b)~\textit{Occlusion:} equipment frequently sits inside or around other equipment, with 7.85\% of instance pairs having overlapping bounding boxes in IndustryCAD (vs.\ 0.46\% in FloorPlanCAD), making it harder to separate adjacent instances. (c)~\textit{Scale variation:} equipment size varies widely, with the ratio between the 99th- and 10th-percentile instance bounding-box areas reaching 510$\times$ (vs.\ 150$\times$ in FloorPlanCAD), so small instances are easily overlooked while large ones contain complex internal structures.

\subsection{Experimental Setup}
\label{sec:experiment_settings}
\paragraph{Implementation Details}
\label{sec:implementation_details}
Our autoregressive transformer uses 6 decoder layers with hidden size $d_{\text{model}}=768$. The continuous transformation attributes (translation, rotation, and scale) are each uniformly discretized into $K=128$ equal-width bins over their valid range, and the model predicts a categorical distribution over bins per attribute. We train the model with AdamW (weight decay 0.05) for 500 epochs on 8 RTX-4090 GPUs, using a base learning rate $5\times 10^{-4}$ with 5\% linear warm-up and cosine decay schedule. We set the per-GPU batch size to 4 with gradient accumulation steps of 4.
Ground-truth foreground instances are ordered left-to-right and then top-to-bottom based on their 2D translations.

\paragraph{Baselines}
\label{sec:baselines}
We compare against three families of approaches:
\begin{itemize}[leftmargin=*]
\item \emph{Symbol spotting.} Primitive-wise methods (PanCADNet~\cite{fan2021floorplancad}, CADTransformer~\cite{fan2022cadtransformer}, GAT-CADNet~\cite{zheng2022gat}) and instance-wise methods (SymPoint~\cite{liu2024symbol}, SymPointV2~\cite{liu2024sympointv2}, CADSpotting~\cite{yang2024cadspotting}, VecFormer~\cite{wei2025point}), all of which formulate spotting as segmentation over input primitives.
\item \emph{Shape abstraction.} 
We reimplement the shape abstraction paradigm by removing the template input branch from our architecture, so that the model decomposes each target into parameterized primitives with implicitly memorized geometry, mirroring the design of PrimitiveAnything~\cite{ye2025primitiveanything}. 
\item \emph{Scene CAD recomposition.} We reimplement the detect-then-fit paradigm~\cite{avetisyan2019scan2cad,kuo2020mask2cad,gumeli2022roca,li2023autoscan2cad}
for our setting using DEIM~\cite{huang2025deim} as the representative detector: DEIM detects instances with bounding boxes and predicts semantics, then templates are retrieved and aligned via ICP~\cite{besl1992method}.
Specifically, since DEIM does not predict the template identity or rotation, we apply ICP to candidate templates within the predicted semantic category using the primitives inside the detected bounding box. The best-aligned candidate is selected as the fitted template, and its
rotation is recovered from the ICP alignment.
\end{itemize}
For both template-fitting baselines, the predicted templates are processed through the same back-projection module as ours.

\paragraph{Evaluation Metrics}
Following~\cite{fan2021floorplancad}, we report Panoptic Quality (PQ) as our primary metric:
$$\text{PQ} = \frac{\sum_{(s_p, s_g) \in \text{TP}} \text{IoU}(s_p, s_g)}{|\text{TP}| + \frac{1}{2} |\text{FP}| + \frac{1}{2} |\text{FN}|},$$
where IoU is length-weighted given the vector nature of primitives.

However, prior works~\cite{liu2024symbol, wei2025point} often adopt a GT-centric greedy matching protocol without a global one-to-one constraint, which allows multiple redundant predictions to be repeatedly matched to the same ground-truth instance and incorrectly counted as TPs, although they should be treated as FPs. This leads to an overestimation of PQ by under-penalizing duplicate predictions.
To ensure a fair and standard evaluation, we enforce strict one-to-one matching by employing the Hungarian algorithm to compute the optimal global assignment between the full sets of predictions and ground truths. We therefore additionally report PQ under this strict Hungarian protocol; detailed evaluation procedures (including the TP/FP/FN counting rules and pseudo-code) are provided in the Supplementary Materials.

\subsection{Results on FloorPlanCAD}
\label{sec:floorplan_results}
\paragraph{Quantitative Evaluation}
Table~\ref{tab:floorplancad} reports results on FloorPlanCAD with both the Greedy and strict Hungarian protocols, where $\text{PQ}_{\text{fg}}$ and $\text{PQ}_{\text{bg}}$ evaluate foreground instances and background regions, respectively.
All methods perform worse under Hungarian matching, with a particularly large gap for VecFormer (91.1 $\rightarrow$ 48.9), confirming that duplicate predictions are strongly penalized under strict one-to-one matching.

\begin{table}[t]
\centering
\caption{Quantitative comparison on the IndustryCAD test set. $\text{PQ}_g$ uses the GT-centric greedy matching protocol from prior work, while $\text{PQ}_h$ enforces a strict global one-to-one assignment via the Hungarian algorithm. The best and second-best results are highlighted by color.}
\label{tab:industrycad}

\begin{tabularx}{\linewidth}{l YY}
\toprule
Methods & $\text{PQ}_g$ & $\text{PQ}_h$ \\
\midrule

SymPoint \cite{liu2024symbol}          & 26.2 & 20.3 \\
SymPointV2 \cite{liu2024sympointv2}    & 34.7 & 28.5 \\
CADSpotting \cite{yang2024cadspotting} & 46.6 & 38.8 \\
VecFormer \cite{wei2025point}          & 66.9 & 44.4 \\
\midrule

PrimitiveAnything \cite{ye2025primitiveanything}      & 64.0 & 53.0 \\
DEIM+ICP \cite{huang2025deim}          & \cellcolor{tabsecond}67.8 & \cellcolor{tabsecond}56.6 \\
\midrule

Ours                                   & \cellcolor{tabfirst}73.9 & \cellcolor{tabfirst}67.0 \\
\bottomrule
\end{tabularx}
\end{table}

Under the Hungarian protocol, our method achieves the best overall PQ (80.1), outperforming the previous best SymPointV2 by 3.1 points. The gain is primarily driven by $\text{PQ}_{\text{fg}}$ (81.7), where our template fitting formulation ensures that every predicted instance is structurally complete by construction, directly eliminating the merge and fragmentation errors that limit segmentation-based methods. For background, our category-specific tokens yield the best $\text{PQ}_{\text{bg}}$ (65.4), as each token is statically bound to a fixed semantic category, avoiding the assignment ambiguity of DETR-style dynamic queries.
Compared with the other template-fitting paradigms, our method outperforms PrimitiveAnything by 8.7 PQ points and DEIM+ICP by 5.7 points. These margins confirm that explicit geometric memory is more effective than implicit template memorization, and that joint retrieve-and-fit avoids the error accumulation of staged detect-then-fit pipelines.

In terms of efficiency, our autoregressive architecture and the inclusion of the template library increase inference latency (279.0~ms vs.\ 32.9~ms for the fastest method); this overhead remains practical for typical scenes in our target domain.


\begin{figure*}[t]
    \centering
    \includegraphics[width=0.98\textwidth]{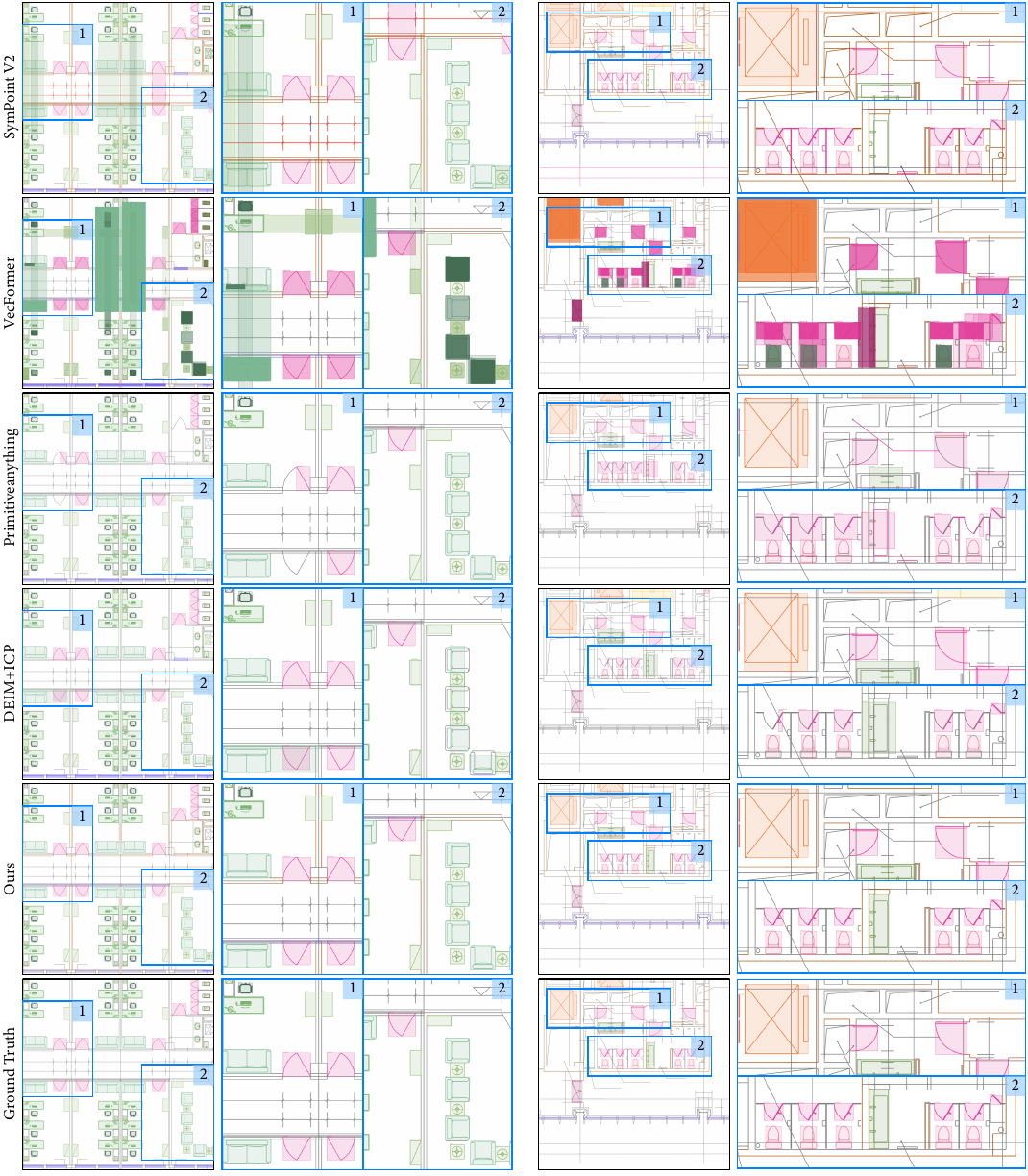}
    \Description{Qualitative comparison on FloorplanCAD. Colors denote predicted categories, boxes mark instances, and blue insets highlight details.}
    \caption{Qualitative comparison on FloorplanCAD. Colors denote predicted categories, boxes mark instances, and blue insets highlight details.}
    \label{fig:qualitative_results}
\end{figure*}

\paragraph{Qualitative Evaluation}
Figure~\ref{fig:qualitative_results} compares our results against representative baselines from two paradigms: Symbol Spotting (SymPointV2, VecFormer) and Template Fitting (Shape Abstraction with PrimitiveAnything, Scene CAD Recomposition with DEIM+ICP).

Symbol Spotting baselines fail at instance grouping. SymPointV2 over-merges spatially distant primitives into a single instance, producing overly large bounding boxes that span across the room, as seen on the left of the first row in Figure~\ref{fig:qualitative_results}. The highlighted elongated box around a chair in the same row further illustrates how it incorrectly includes surrounding wall lines in the foreground prediction, while its background predictions drift visibly against the ground truth. In the second row, VecFormer stacks multiple overlapping boxes for the same object and similarly merges distinct objects into a single instance.

The Template Fitting baselines avoid this over-merging by construction: every predicted instance is an instantiation of a valid template, so its constituent primitives automatically form a structurally coherent group. However, both baselines exhibit their own failure modes. PrimitiveAnything memorizes template geometry implicitly within network weights, so its predicted transformation parameters often drift, yielding instances that do not tightly cover the underlying primitives, as seen in the highlighted red door in the third row. Scene CAD Recomposition (DEIM+ICP) inherits the limitation of its detection stage: when the detected bounding box is misaligned in position or scale, as in the over-extended box around the green sofa in the fourth row, the subsequent ICP fitting cannot recover, producing misplaced or wrongly-scaled templates.

In contrast, our method produces cleaner and more precise results. By grounding predictions in explicit geometric memory from the template library and unifying retrieval with pose prediction end-to-end, it avoids both the parameter drift of implicit memorization and the cascading errors of staged detection, while maintaining accurate background segmentation that closely matches the ground truth.

\begin{table}[t]
\centering
\caption{$\text{PQ}_h$ on IndustryCAD across inter-instance overlap bins. Overlap degree is the percentage of each instance covered by others; $B_1$--$B_5$ denote $[0,0.05)$, $[0.05,0.25)$, $[0.25,0.50)$, $[0.50,0.75)$, and $[0.75,1.00]$.}
\label{tab:overlap}

\setlength{\tabcolsep}{4pt}
\renewcommand{\arraystretch}{1.1}

\begin{tabularx}{\columnwidth}{l *{5}{Y}}
\toprule
\multirow{2}{*}{Methods}
& \multicolumn{5}{c}{Overlap degree} \\
\cmidrule(lr){2-6}
& \leavevmode$B_1$ & \leavevmode$B_2$ & \leavevmode$B_3$ & \leavevmode$B_4$ & \leavevmode$B_5$ \\
\midrule
DEIM+ICP~\cite{huang2025deim} & \leavevmode 76.4 & \leavevmode 67.4 & \leavevmode 55.3 & \leavevmode 37.8 & \leavevmode 26.9 \\
Ours & \leavevmode 83.4 & \leavevmode 71.0 & \leavevmode 59.0 & \leavevmode 44.7 & \leavevmode 54.4 \\
\bottomrule
\end{tabularx}
\end{table}

\subsection{Results on IndustryCAD}
\label{sec:industrycad_results}

\paragraph{Quantitative Evaluation}
Table~\ref{tab:industrycad} reports the comparison on IndustryCAD. All methods degrade from FloorPlanCAD to IndustryCAD, but the drop differs sharply across paradigms. Symbol Spotting collapses, with SymPointV2 falling from $77.0$ to $28.5$ PQ under the Hungarian protocol; in contrast, the three Template Fitting baselines (PrimitiveAnything, DEIM+ICP, and Ours) all stay in the $53.0$ to $67.0$ range, well above any Symbol Spotting method, and our method drops the least ($80.1 \rightarrow 67.0$). This indicates that the Template Fitting formulation is more robust under heavy complexity, occlusion, and scale variation.

Within the Template Fitting paradigm, DEIM+ICP outperforms PrimitiveAnything by $3.0$ / $3.6$ PQ on the two benchmarks, indicating that explicit template input consistently beats implicit memorization. Our method further outperforms DEIM+ICP by $5.7$ / $10.4$ PQ, and the gap widens on the harder benchmark, validating our joint retrieve-and-fit over the staged detect-then-fit pipeline.

To further analyze the larger performance gap between TempFit and the DEIM+ICP (detect-then-fit) baseline on IndustryCAD, we compare their performance under different degrees of inter-instance overlap. We define the overlap degree as the percentage of an instance covered by other instances and report PQ using Hungarian matching across five overlap intervals in Table~\ref{tab:overlap}. TempFit achieves higher PQ than DEIM+ICP in every interval, with the largest margin in the most heavily overlapped group (54.4 vs.\ 26.9). The key difference lies in how the fitted templates are obtained. DEIM+ICP aligns candidate templates to the primitives inside each detected bounding box; when instances overlap, the box contains foreign primitives that can corrupt both template selection and pose alignment. TempFit instead predicts the template identity and pose from the full drawing context without relying on a potentially contaminated local primitive set.

\begin{figure*}[!t]
    \centering
    \includegraphics[width=0.98\textwidth]{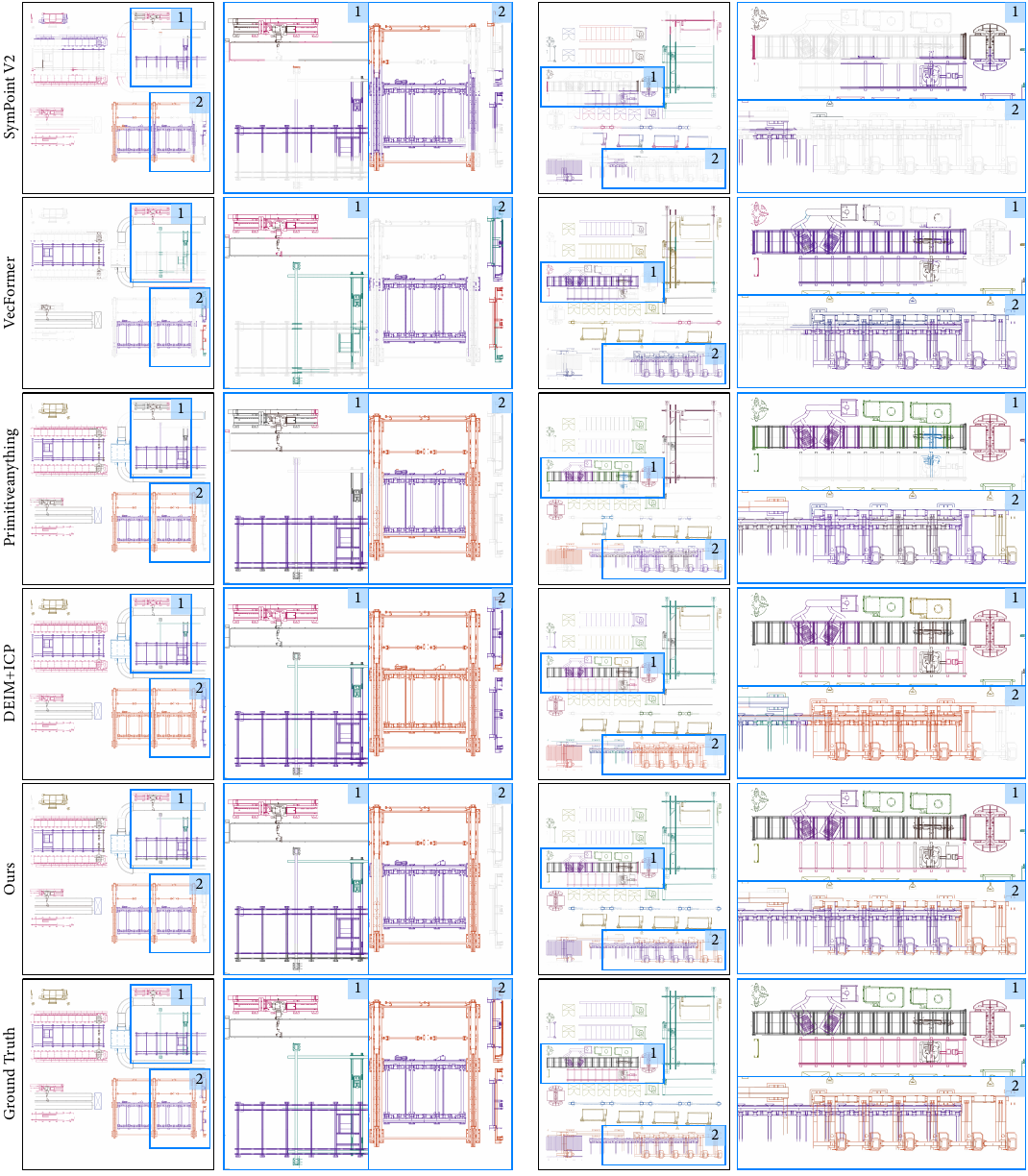}
    \Description{Qualitative comparison on IndustryCAD. Colors denote predicted categories, and blue insets highlight details.}
    \caption{Qualitative comparison on IndustryCAD. Colors denote predicted categories, and blue insets highlight details.}
    \label{fig:qualitative_industry}
\end{figure*}

\paragraph{Qualitative Evaluation}
Figure~\ref{fig:qualitative_industry} compares results against representative baselines on IndustryCAD.
The higher per-instance complexity and heavier inter-instance overlap of IndustryCAD amplify the failure modes of every baseline.

Symbol Spotting baselines fail in two ways. First, they struggle to recover complex instances as a whole, often breaking a single instance into several disjoint regions or missing it entirely. Second, they degrade in heavily overlapping regions, where primitives from different instances share the same drawing area. These failures are visible in the first row of Figure~\ref{fig:qualitative_industry}: the highlighted brown shelf on the left is broken into several disjoint regions, while highlight area on the right reveals both large gray areas where entire instances are missed and overlapping regions where primitives from different instances are not separated.

Template Fitting baselines mitigate the missing-instance problem to some extent, since every prediction is bound to a valid template. However, they still exhibit their own failure modes. PrimitiveAnything, which lacks explicit template input, struggles to fully reconstruct the geometry of complex industrial instances; both the green shelf in the left highlight and the conveyor below the right highlight are only partially recovered. DEIM+ICP, which localizes instances through detection-stage bounding boxes, cannot precisely separate distinct instances in heavily overlapping regions, instead merging multiple instances into a single prediction within the left highlight.

In contrast, our method produces cleaner and more precise results. By grounding predictions in explicit geometric memory and unifying retrieval with pose prediction end-to-end, it better preserves the geometry of each complex symbol and separates distinct instances even in heavily overlapping regions.

\begin{figure*}[!t]
    \centering
    \includegraphics[width=\linewidth]{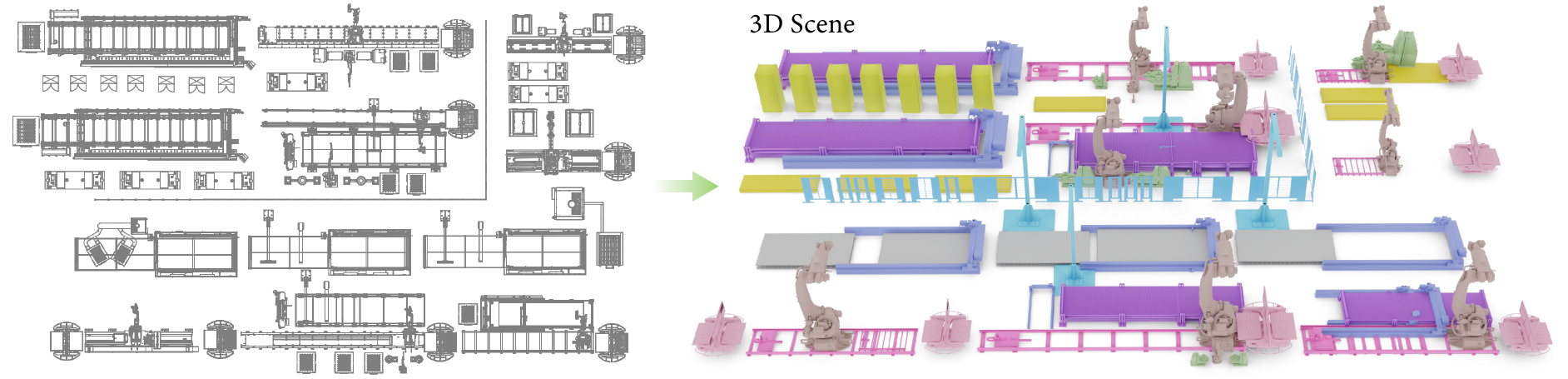}
    \Description{Predicted (template, transformation) pairs on IndustryCAD can be lifted back to 3D and directly assembled into a 3D industrial scene.}
    \caption{Predicted (template, transformation) pairs on IndustryCAD can be lifted back to 3D and directly assembled into a 3D industrial scene.}
    \label{fig:application}
\end{figure*}

Since the 2D templates in IndustryCAD are projected from 3D asset library at the data source, the predicted (template, transformation) pairs can be lifted back to 3D and directly assembled into the underlying 3D industrial scene, as shown in Figure~\ref{fig:application}.

\subsection{Ablation Studies}
\label{sec:ablation}
Table~\ref{tab:ablation} summarizes the ablation results on the FloorPlanCAD test set, including the contributions of the core design choices and the effect of instance ordering.

\paragraph{Effect of Template Priors}
To validate the effectiveness of the explicit template memories, we remove the template tokens and employ a standard linear head for direct class prediction. 
As shown in Row 1 of Table \ref{tab:ablation}, this leads to a significant performance drop, particularly in $\text{PQ}_{\text{fg}}$. This confirms that explicitly encoding object templates provides essential geometric guidance, enabling the model to better distinguish symbols
in complex drawings.

\paragraph{Effect of Category-Specific Background Tokens}
We replace our background tokens with the same number of standard DETR-style queries. These queries are category-agnostic and dynamically assigned to the background-mask targets through bipartite matching. As shown in Row 2, $\text{PQ}_{\text{bg}}$ drops significantly by 6.1 points, confirming that static binding eliminates the instability inherent in dynamic assignment. Notably, this degradation also negatively impacts $\text{PQ}_{\text{fg}}$, suggesting that a stable and precise background layout serves as a critical contextual constraint for foreground object localization.

\paragraph{Effect of Robust Assignment}
We replaced our proposed primitive assignment during back projection with a heuristic Euclidean distance strategy (assigning primitives to the spatially closest template). The results in Row 3 demonstrate that spatial proximity alone is insufficient, since nearby primitives often belong to different instances in cluttered regions.

\paragraph{Effect of Instance Ordering}
To evaluate the model's sensitivity to the predefined instance order, we retrain a variant with target sequences ordered by a Hilbert curve computed from the same 2D instance translations, with all other settings unchanged. As shown in Row 4 of Table~\ref{tab:ablation}, its PQ decreases only slightly from 80.1 to 79.2. The change mainly occurs in foreground prediction, where $\mathrm{PQ}_{\mathrm{fg}}$ decreases by 1.2 points, while $\mathrm{PQ}_{\mathrm{bg}}$ remains nearly unchanged (65.4 vs.\ 65.1). These results indicate that TempFit is robust to different spatially coherent instance orderings.

\begin{table}[t]
\centering
\caption{Ablation studies on the FloorPlanCAD test set. The first three variants assess the core design choices, while the Hilbert-curve variant evaluates the effect of instance ordering.}
\label{tab:ablation}

\setlength{\tabcolsep}{4pt}
\renewcommand{\arraystretch}{1.05}

\begin{tabularx}{\columnwidth}{l *{3}{Y}}
\toprule
Methods
& PQ
& $\mathrm{PQ}_{\mathrm{fg}}$
& $\mathrm{PQ}_{\mathrm{bg}}$ \\
\midrule

w/o Template Memory
&71.4 & 73.9 & 61.7 \\

DETR-style Background Queries
&74.7  &76.2 & 59.3 \\

w/o Robust Assignment
&75.9 &77.7 & 63.3 \\

Hilbert-curve Ordering
&\leavevmode 79.2 &\leavevmode 80.5 &\leavevmode 65.1 \\

Ours (Full Model)
& \cellcolor{tabfirst}80.1 & \cellcolor{tabfirst}81.7 & \cellcolor{tabfirst}65.4 \\

\bottomrule
\end{tabularx}
\end{table}

\subsection{Robustness Analysis}
\label{sec:robustness}

\paragraph{Analysis of Autoregressive Error Accumulation}
On the FloorPlanCAD test set, we analyze how an incorrect prediction affects later instances. Predictions are compared with ground truth at the same positions in the predefined spatial order. A prediction is correct if its retrieved template identity matches the corresponding ground truth and their IoU exceeds 0.5. For each sequence with an error, we locate the first incorrect prediction and measure correctness at the next four positions ($+1$ to $+4$).

The prediction immediately after the first error is the most affected, with 78.6\% correctness. Correctness then increases to 84.3\%, 85.0\%, and 87.6\% at positions $+2$, $+3$, and $+4$, respectively, indicating that the influence of an incorrect prediction diminishes rather than propagating persistently through the sequence. This recovery occurs because, despite attending to previous outputs through self-attention, each decoding step remains grounded in the full encoded drawing through cross-attention.

\paragraph{Robustness to Template Library Size}
We analyze the robustness of our method by randomly reducing the size of the template library during inference. As shown in Table~\ref{tab:robustness}, the performance degrades gracefully rather than collapsing, even as the library becomes sparse. This demonstrates the model's strong generalization capability: even when the exact ground-truth template is missing, the model successfully retrieves a semantically similar alternative from the remaining library and transforms it to fit the target instance.

\begin{table}[t]
\centering
\caption{Robustness analysis on template library size. We evaluate the robustness of our method by randomly reducing the size of the template library during inference.}
\label{tab:robustness}

\setlength{\tabcolsep}{6pt}
\renewcommand{\arraystretch}{1.1}

\begin{tabularx}{\columnwidth}{l *{3}{Y}}
\toprule
Keep Ratio
& PQ
& $\mathrm{PQ}_{\mathrm{fg}}$
& $\mathrm{PQ}_{\mathrm{bg}}$ \\
\midrule

25\%
& 65.2 & 67.2 & 58.9 \\

50\%
& 72.0 & 74.3 & 61.8 \\

75\%
& 76.1 & 78.4 & 63.2 \\

100\% (Full Model)
& \cellcolor{tabfirst}80.1
& \cellcolor{tabfirst}81.7
& \cellcolor{tabfirst}65.4
\\
\bottomrule
\end{tabularx}
\end{table}

\paragraph{Zero-Shot Cross-Dataset Evaluation}
As a stress test, we evaluate our method on SESYD~\cite{delalandre2010generation} without fine-tuning. As shown in Table~\ref{tab:sesyd}, our method achieves PQ~$24.67$, outperforming the strongest baseline SymPointV2 ($16.23$) by a large margin, as well as the template-fitting methods DEIM+ICP ($11.77$) and PrimitiveAnything ($9.36$), and VecFormer ($8.43$). This robustness stems from the geometric nature of our retrieve-and-fit mechanism. Rather than directly predicting instance labels from learned features, which are tightly coupled to the training distribution, our model retrieves the most geometrically similar template from the library and adapts it via an explicit 2D transformation. Even when no template exactly matches the target symbol, a sufficiently similar template can be fitted with an appropriate pose, yielding a reasonable result. This geometric grounding makes our method more tolerant to distribution shift than segmentation-based symbol spotting, whose features are distribution-specific. The other template-fitting baselines, despite also fitting templates, are less robust to domain shift for different reasons: PrimitiveAnything memorizes template geometry implicitly in network weights, which remain tied to the training distribution, while DEIM+ICP relies on a detection stage whose learned features equally suffer from domain shift, causing ICP to start from poorly localized boxes.

\begin{table}[t]
\centering
\caption{Zero-shot cross-dataset evaluation on SESYD. Our method outperforms all baselines by a clear margin despite the substantial geometric domain gap from FloorPlanCAD.}
\label{tab:sesyd}

\setlength{\tabcolsep}{4pt}
\renewcommand{\arraystretch}{1.05}

\begin{tabularx}{\columnwidth}{l Y}
\toprule
Methods & PQ \\
\midrule

VecFormer~\cite{wei2025point}
&  8.43 \\

SymPointV2~\cite{liu2024sympointv2}
& 16.23 \\
\midrule

PrimitiveAnything~\cite{ye2025primitiveanything}
&  9.36 \\

DEIM+ICP~\cite{huang2025deim}
& 11.77 \\
\midrule

Ours
& \cellcolor{tabfirst}24.67 \\

\bottomrule
\end{tabularx}
\end{table}

\section{Conclusion}
This work presents a new paradigm for Panoptic Symbol Spotting in CAD drawings. We reformulate the task as a \emph{Template Fitting} problem: foreground instances are obtained by retrieving canonical shapes from a template library and aligning them to the drawing, which explicitly enforces structural completeness and geometric integrity. Based on this formulation, we introduce \emph{TempFit}, an autoregressive transformer that treats the template library as explicit geometric memory and unifies template retrieval with pose prediction end-to-end. It parses background regions in parallel via learnable, category-specific tokens to capture the global layout, and synthesizes foreground instances through an autoregressive fitting process that retrieves templates and predicts transformation attributes. Extensive experiments on FloorplanCAD and our newly collected IndustryCAD demonstrate state-of-the-art performance, with particularly large gains on the more challenging IndustryCAD, validating our design choices over both Symbol Spotting and Template Fitting baselines.

\begin{figure}[!t]
    \centering
    \makebox[\linewidth][c]{%
    \begin{minipage}{0.49\linewidth}
        \centering
        \includegraphics[width=\linewidth]{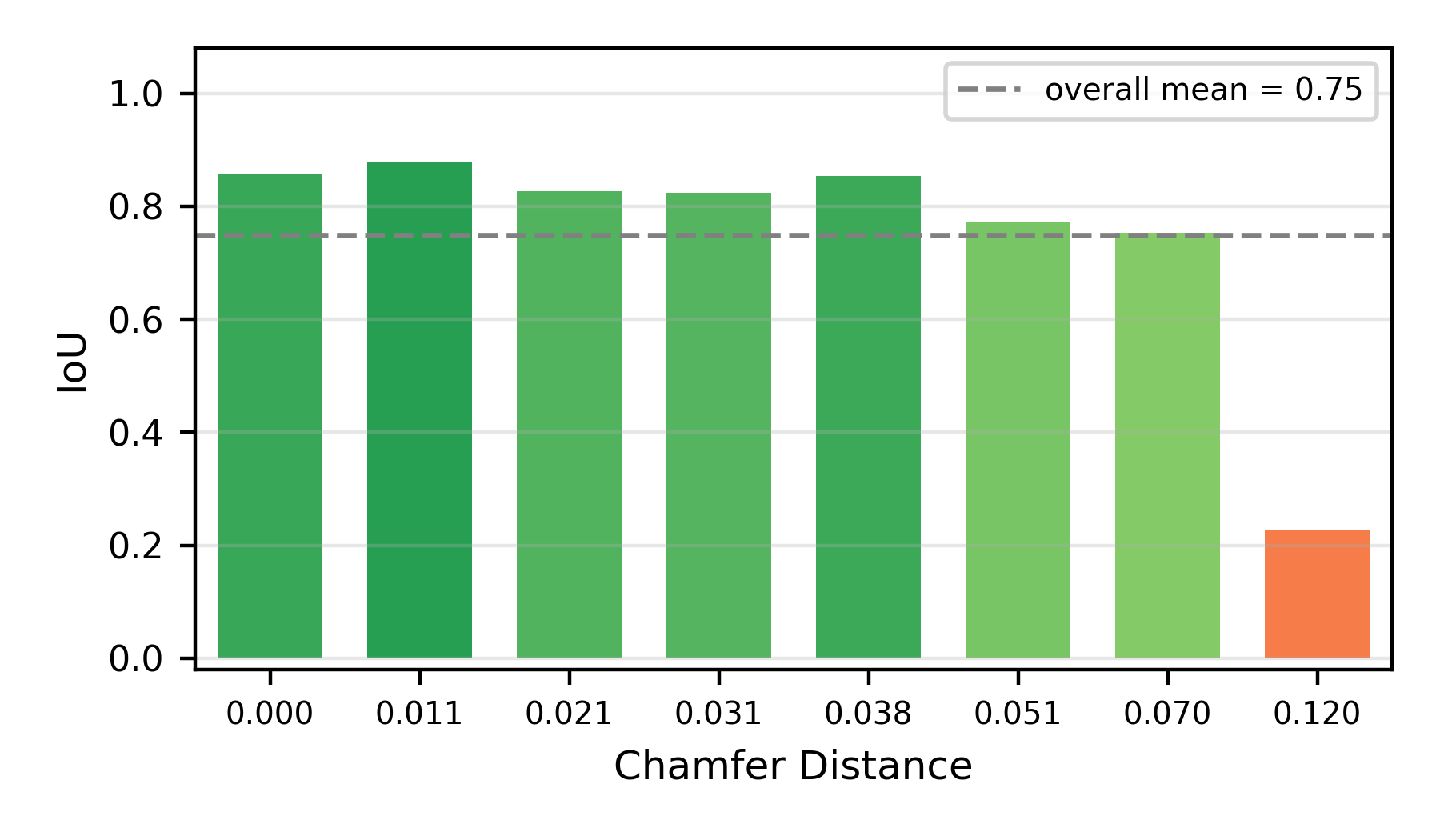}\\
        \small (a) IoU vs.\ template distance.
    \end{minipage}\hspace{0pt}%
    \begin{minipage}{0.49\linewidth}
        \centering
        \includegraphics[width=\linewidth]{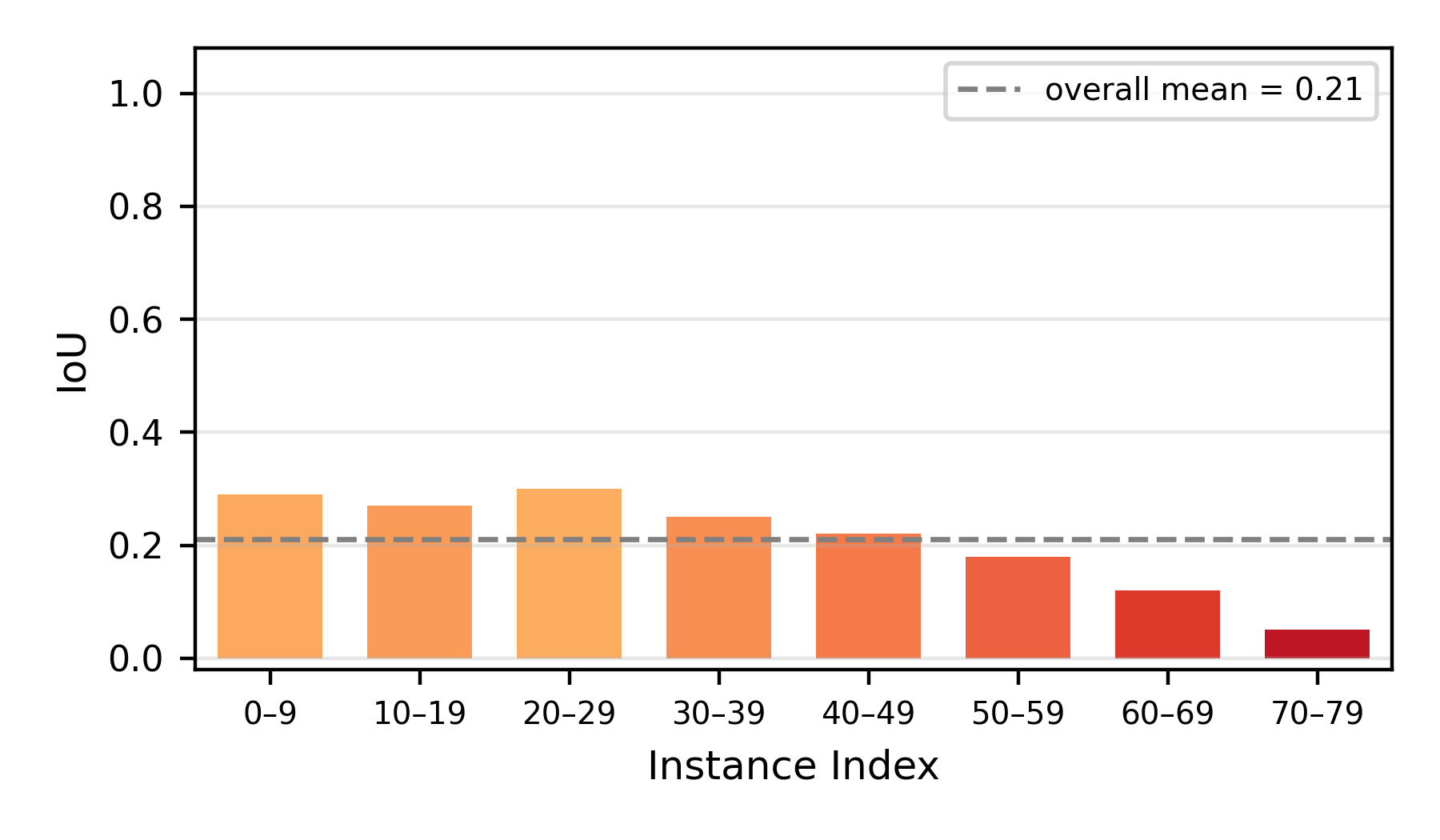}\\
        \small (b) IoU vs.\ decoding position.
    \end{minipage}}
    \Description{Failure case analysis. (a) Mean IoU between the predicted and ground-truth instances drops once the Chamfer distance from the ground-truth instance to its nearest template exceeds $\sim\!0.07$. (b) For a representative drawing with $n_{\text{inst}}=80$, the bin-averaged IoU between predicted and ground-truth instances decays along the autoregressive decoding position.}
    \caption{Failure case analysis. (a) Mean IoU between the predicted and ground-truth instances drops once the Chamfer distance from the ground-truth instance to its nearest template exceeds $\sim\!0.07$. (b) For a representative drawing with $n_{\text{inst}}=80$, the bin-averaged IoU between predicted and ground-truth instances decays along the autoregressive decoding position.}
    \label{fig:failure_cases}
\end{figure}

\paragraph{Limitations and Future Work}
Our method exhibits two failure modes, analyzed quantitatively in Figure~\ref{fig:failure_cases}.
First, fitting quality degrades when an instance is geometrically far from every template in the library. As shown in Figure~\ref{fig:failure_cases}(a), the IoU of matched predictions stays roughly stable at small to moderate Chamfer distances but drops sharply once the distance exceeds $\sim\!0.07$, indicating that instances falling outside the geometric coverage of the library cannot be faithfully recovered.
Second, our method can fail on drawings with exceptionally many instances, which are rare in the training data and introduce a substantial scale shift.
The failure case in Figure~\ref{fig:failure_cases}(b) involves an unusually long sequence of 80 instances, a sequence length observed in less than 1\% of the training data, whereas more than 80\% of the training blocks contain fewer than 25 instances. With so many instances in a single block, each instance appears at a substantially smaller scale than those observed during training, creating an out-of-distribution condition. Under this scale shift, the model already fails at the first decoding step, with an IoU of only $\sim\!0.28$, followed by progressive degradation along the unusually long autoregressive decoding chain, with IoU falling below $0.05$ toward the end of the sequence.
The token-by-token generation also limits inference speed on large drawings.
Future work includes extending the template library or exploring open-vocabulary template adaptation to handle out-of-library cases, and predicting all attributes of an instance jointly to shorten the decoding chain and accelerate inference.

\begin{acks}
We thank the anonymous reviewers for their valuable feedback and constructive suggestions, which helped improve this work.
This work is supported by the National Key R\&D Program of China (2024YFB2809102), the National Natural Science Foundation of China (62322207, 62325211, 62132021), the Research Fund of Jiangsu Key Laboratory of AI for Industries (E6420016G8), and Manfred Lau's start-up funds from Lakehead University (Faculty of Science \& Environmental Studies).
\end{acks}

\bibliographystyle{ACM-Reference-Format}
\bibliography{reference}



\end{document}